# Co-Designing a wiki-based community knowledge management system for personal science

Katharina Kloppenborg[1], Mad Price Ball[2], Steven Jonas[3], Gary Isaac Wolf[3], Bastian Greshake Tzovaras[1,3,4]*

[1] Inserm U1284, Université Paris Cité, Paris, France
[2] Open Humans Foundation, Sanford, NC, USA
[3] Quantified Self Labs, Berkeley, CA, USA
[4] The Alan Turing Institute, London, United Kingdom

* bgreshake@googlemail.com

## Abstract

Personal science is the practice of addressing personally relevant health questions through self-research. Implementing personal science can be challenging, due to the need to develop and adopt research protocols, tools, and methods. While online communities can provide valuable peer support, tools for systematically accessing community knowledge are lacking. The objective of this study is to apply a participatory design process involving a community of personal science practitioners to develop a peer-produced knowledge base that supports the needs of practitioners as consumers and contributors of knowledge. The process led to the development of the Personal Science Wiki, an open repository for documenting and accessing individual self-tracking projects while facilitating the establishment of consensus knowledge. After initial design iterations and a field testing phase, we performed a user study with 21 participants to test and improve the platform, and to explore suitable information architectures. The study deepened our understanding of barriers to scaling the personal science community, established an infrastructure for knowledge management actively used by the community, and provided lessons on challenges, information needs, representations, and architectures to support individuals with their personal health inquiries.

## Keywords

# Introduction

Patient-led research has gained recognition for its innovative potential in addressing unexplored research areas identified by patients [1–4]. A specific form of participant-led research is personal science, in which individuals employ empirical methods to explore health questions relevant to them [5,6]. In these self-directed n-of-1 studies [7], practitioners use commercially available or self-made tracking devices, design protocols, and choose analysis tools to interpret their health data. As individuals perform all or most of these research steps outside traditional academic research settings, personal science can be interpreted as a form of citizen science [8–10]. Personal science can yield unique insights for individuals [11,12] and enhance their sense of agency and quality of life [13]. Moreover, self-research can drive the development of new approaches or tools [14,15], inspiring clinical research with these tools and ideas [16].

While there is no epistemic necessity to collaborate, communities nevertheless play a crucial role in personal science. Practitioners often share experiences and feedback through various channels such as online and offline meetings, forums, blogs, chats, or social media [17]. Two notable communities dedicated to personal science are Quantified Self, which has pioneered the concept of personal science [5], and Open Humans. Since 2008, Quantified Self has hosted numerous online and offline "Show and Tell" events in which self-researchers present their projects under the slogan "self-knowledge through numbers" [18]. They maintain an active online forum and an archive of meeting recordings [5]. Quantified Self has received extensive media coverage [19], and it has been a subject of research involving health care, personal informatics [20], and its role as a cultural phenomenon [17]. The Open Humans Foundation, established in 2015 as an extension of the Harvard Personal Genomes Project [21], aims to "empower individuals and communities around their personal health data for education, health, and research purposes". They manage the Open Humans platform for personal data exploration, participant-led research, and citizen science [22], along with an active Slack community and weekly online community meetings, known as "self-research chats."

Personal science can be intricate, as practitioners typically lead all stages of their research projects themselves and frequently need to tailor protocols, data collection, and analysis tools to their specific needs and contexts [5]. Peer support serves a vital role in providing assistance. Additionally, being part of a community of peers with shared goals and values, exchanging experiences, and learning from one another are key motivators for sustained engagement in self-research practice [23]. Despite active communities, as well as the acknowledged importance of peer support, there is an observed tendency for individuals to create their methods from scratch when they start a new project, imposing barriers to the scaling of the practice to a wider audience. As a possible reason, lacking systematic access to community knowledge has been evoked [5]. To date, in spite of existing pathways for documentation and exchange, no accumulation of a shared community knowledge base has been observed [9,16].

## Objective

This study seeks to explore and document how a community-based co-creation process was used to address knowledge management challenges in personal science. As part of this work, we have implemented a functional knowledge management infrastructure, tested and improved it based on feedback through active use by community members, and performed usability tests as well as a card sorting study to compare practitioners' mental models with our existing information architecture.

# Methods

In this study, we employed a mixed methods approach to co-create an infrastructure for a shared knowledge management resource for personal science. First, we engaged in a participatory design approach with an existing community of practice to identify barriers related to knowledge sharing and community support in personal science. Based on these insights, we iteratively designed and discussed potential solutions, including the development, testing, and improvement of a functional prototype. Following the design phase, we conducted a user study to assess and improve the usability of the prototype, as well as to gain insights into user needs regarding information architecture knowledge related to personal science, using a card sorting method.

## Design Process

This part of the study makes use of the design thinking methodology, which serves to support matching users' needs with technical feasibility and strategic goals [24]. More specifically, we used the "double diamond" design process model [25]. This model aims to lead from a general problem statement to a specific solution, via two phases, or "diamonds" (see Figure 1). The first phase deals with exploring the "problem space", in the sense of identifying a list of problems through user research, and then deciding on a specific problem to solve. The second phase is about the "solution space", prototyping and user-testing a range of potential solutions, until converging on one solution that will be further developed.

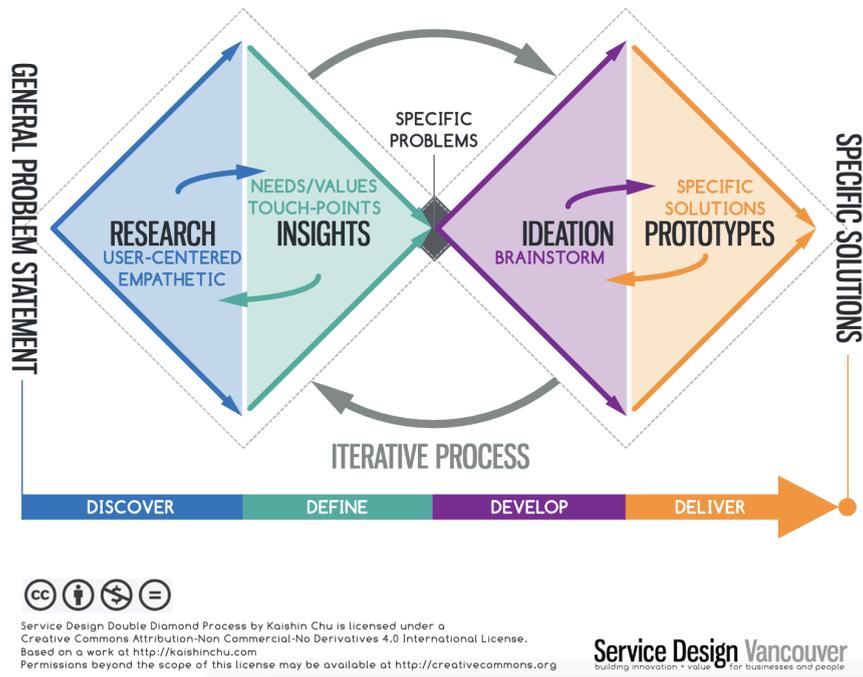

*Figure 1 The double diamond design process. Image from [26]*

Addressing the problem space involved identifying stakeholders, considering who could be involved in the design process, and developing and prioritising user personas. Personas, which represent archetypical user types with factors like demographics, needs, wishes, and challenges, were created to serve as mnemonic devices, making user needs prominent from the start of the design process [27]. Following this, KK extracted excerpts regarding barriers, frustrations, motivations, and ideas from interviews with self-tracking practitioners previously conducted by members of our research team [23]. These excerpts were visualised on a collaboration board using Miro [28], clustered according to topic, and categorised using the "rose bud thorn" method [29]. Different colours were assigned to comments by interviewees, representing their motivations, frustrations, and ideas or areas of potential. The outcomes were then discussed with BGT and MPB, who are also community managers of Open Humans, to identify the most relevant and promising elements.

Based on these results, a list of potential solution approaches was brainstormed and discussed, leading to the development of a platform prototype mockup using the browser-based design tool Figma [30]. The prototype's advantages and disadvantages were discussed, leading to another iteration, this time incorporating a wiki approach. After further discussions, presentations to community stakeholders, and a community meeting, the wiki-based approach was selected for further development. This solution, the "Personal Science Wiki" was deployed and made accessible for the community to test and use.

# User study

To assess the usability of the Personal Science Wiki, learn about user needs, and to make use of the newly accumulated content to learn whether the information structures of the wiki are in line with mental models of practitioners, we conducted a mixed-methods user study. The study included a usability test, a card sorting task, and a short interview, all conducted online with one participant at a time using video conferencing software. Ethical approval for this user testing was obtained from the Institutional Review Board of the French National Institute of Health and Medical Research (Inserm, IRB00003888) on January 10, 2023. Calls for participation were disseminated through the Quantified Self forum, the Open Humans slack channel, our team's institute's channels, direct email invitations to community members, and snowball sampling.

## Usability test

To evaluate the usability [31] of the wiki, we developed a test protocol encompassing common user tasks. As there isn't a single specific workflow for effectively using the Personal Science Wiki, we compiled a list of perceived common use-cases. The study protocol incorporated four information-seeking tasks from this list, covering typical self-tracking topics across the wiki's main categories and using semantic linking functionality. In addition, general feedback questions and two open search tasks, where participants freely explored the wiki to gather information for self-research projects, were included. The primary tasks are outlined in Table 1.

| Item No. | Task / Question |
|---|---|
| 1 | "Please look for a device that records sleep data." |
| 2 | "Please look for projects other people have done related to sleep tracking." |
| 3 | "Please look for people who have worked on activity tracking." |
| 4 | "Please look for projects that use a Fitbit device." |
| 5 | "Imagine you would like to improve your sleep, and stumbled across this wiki. Please explore the wiki in order to see if and how it could help you to implement your project." |
| 6 | "Based on experience/interests of participant: Imagine you would like to do a project about condition/topic X and you stumbled across this wiki. How would you try to find information that might be useful for you?" |

*Table 1 Tasks of the usability tests*

During the study sessions, KK acted as the moderator to guide participants through the tasks, during which participants were instructed to open the Personal Science Wiki homepage and share their screen. The moderator guided them through the search and exploration tasks sequentially, prompting them to verbalise their thoughts throughout the process. The datasets generated from these usability tests consist of transcripts from video recordings of the sessions. These transcripts were initially auto-transcribed using the tactiq browser extension [32], manually corrected, and then anonymized. For each transcript, task-specific summaries were created, detailing information on task completion, ease of completion, difficulties faced, observations made, and comments provided. Utilising these summaries, we compiled a list of usability issues, assigned priorities, and made corresponding adaptations to the wiki. To assess the impact of these changes, the same usability protocol and analysis were repeated with new participants, and the results from both test iterations were compared.

## Card sorting

As part of the study, a card sorting task was conducted to explore if our information architecture aligns with participants' mental models of how they internally organise information related to personal science. Originating from psychology [33], the term "mental models" refers to "personal, internal representations of external reality that people use to interact with the world around them". Also framed as "naive theory", they are understood as sets of causal explanations and organisations of complex systems people use in daily life [34]. They are not necessarily accurate, are unique to each person, and can evolve with new experiences or learnings [35]. Mental models play a role in the development of knowledge management systems, since the structure of the system should resemble users' mental models, so they can find information and thus efficiently use the system. Finding a suitable structure is challenging, because mental models might differ between users, and because of the evolving nature of knowledge within the system [36]. A mismatch between users' mental models and the system architecture can cause usability issues [37].

Card sorting provides a common user research method to elicit mental models and to test or generate information architectures for online resources [38]. We used the *open* card sorting method, in which participants were given virtual "cards" containing the title of content pages along with the lead text. Participants were then asked to group these cards into clusters that make sense to them and to find an overarching label for each cluster. The results help to understand relationships and organisational structures users apply to the given content [35]. To prepare the task, we selected a subset of page titles from the Personal Science Wiki as cards. We aimed to include cards representative of the content, covering the main content categories, as well as cards that were challenging to sort within the information architecture in use. The initial list of 60 cards was reduced to 45 after a test run with a volunteer in order to prevent participant fatigue. A selection of exemplary cards is shown in Table 2. For the full list, please consult the supplementary materials.

| Card | Label |
| --- | --- |
| A Decade of Tracking Headaches | A Decade Of Tracking Headaches is a Show & Tell talk by Stephen Maher [...].The talk was given on 2018/09/22 and is about Pills intake, Sleep, and Stress. |
| Activity tracking | Activity tracking typically describes the act of tracking physical activity[1] and is frequently measured through metrics such as steps, calories burned, distance walked/run, heart rate and [...] |
| Autoethnography using one button tracker and Jupyter notebook | This page provides a step-by-step guide on how one can use a one-button tracker such as the Puck.js to do an autoethnography that combines qualitative and quantitative data. |
| Bangle.js | The Bangle.js[1] is the name of a series of open source smartwatches that are made by Espruino under the leadership by Gordon Williams, who also designed the Puck.js open source hardware that can[...] |
| Blood glucose tracking | Blood glucose tracking involves methods and tools to measure blood glucose levels, commonly in the context of diabetes but also by non-diabetic users. |

*Table 2 A selection of cards used for the card sorting exercise*

Data was collected using the browser-based tool kardSort [39]. Similar to the usability test, a moderator was present via video conferencing, and participants shared their screen while voicing their thoughts out loud.

Both quantitative and qualitative methods were employed for data analysis. Quantitative approaches, combine and analyse data from all participants to find an "average" organisation of items [40]. On the other hand, qualitative approaches provide insights into the reasoning and motivation behind participants' categorizations, and account for individual variations [41]. For the quantitative analysis, we downloaded tabular outputs from kardSort, and conducted hierarchical cluster analysis with average linkage, a commonly used approach recognized for yielding balanced, easily interpretable clusters [38,42]. This method calculates pairwise distance between all cards, and merges them into clusters from the bottom-up until all cards form one cluster. If all participants sorted two cards together, these cards are assigned the lowest distance, and inversely, if all participants sorted two given cards in separate categories, they are assigned the highest distance. Our code for this analysis has been packaged into the open source Python package "cardsort" [43]. For the qualitative analysis, similar to the usability tests, transcripts from video recordings were used. Excerpts from the cleaned and anonymized transcripts were manually tagged using the software taguette [44]. We created tags for each card, categories of feedback, and user-generated cluster labels.

# Results

## Design Process

### Stakeholders and personas

Stakeholders were explored in two ways: Firstly, we sought to identify real community members who could potentially provide valuable input or actively participate in the design process. Secondly, we created archetypical user personas to understand typical motivations and challenges of self-researchers. We compiled lists of active community members known to us, categorising them based on their level of involvement. Beyond the two long-term Open Humans community managers (BGT and MBP), this also included two long-term community managers of Quantified Self (GIW and SJ), who subsequently became more involved in this study. Additionally, we identified other stakeholders closely connected to the community, such as individuals who had published academic articles about personal science, actively participated in community events, or frequently engaged in dedicated forums or Slack channels. These stakeholders were identified for occasional involvement in the design process. In addition to these closely connected stakeholders, we defined the larger community, encompassing individuals practising self-research. These community members had intermittent interactions with the online community, such as sharing results, projects, or questions. We planned to interview or involve them periodically in testing prototypes and gathering feedback.

To create personas, we engaged in a series of design iterations. Initially, several team members independently created persona drafts based on interviews with real community members [23]. From these drafts, common dimensions were extracted, and lists of personas considered relevant by different team members were compiled and refined. This resulted in a final list of 13 personas, with one example provided in Table 3. For the complete list, please refer to the supplementary materials.

| 1 | **Taylor the Techie** |
|---|---|
| 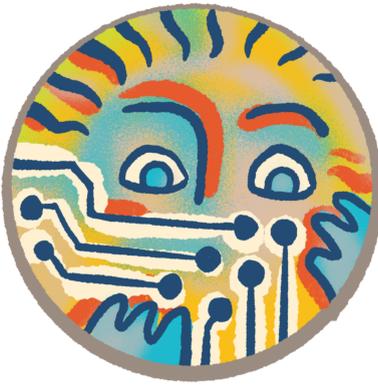 | *"What can I do?"*<br><br>Taylor is a software engineer who is interested in playing with technology at large, including playing around with hardware. Taylor loves collecting data, even if they are not 100% sure what it could be used for.<br><br>**Motivation:** Internally motivated, excited about technology, likes to spend time on it in their freetime<br><br>**Challenges:** Finding a purpose to apply their skills and interests<br><br>**Community:** Self-research chats, likes to help people with their technology- and data-related questions<br><br>**Skills: meta research:** beginner; **data analysis:** intermediate; **coding:** expert |

*Table 3 An exemplary persona representing an archetypal user type in the personal science community*

We divided the persona list into two parts: one consisting of personas to prioritise in the upcoming design process, and one of those not prioritised. The prioritised personas represented individuals that were willing to engage in the complex process of self-research. These individuals were driven by intrinsic motivation and enjoyment of working with data and technology, including experimentation and trial-and-error, and a desire to enhance their health or well-being. On the other hand, personas not prioritised included those who lacked personal interest in self-research in general, or beyond using ready-to-use tools for casual tracking. Additionally, personas wanting to donate their data for others to use or seeking to utilise data donations for larger research studies were deprioritized.

## Exploring the problem space

For our user research, we revisited the existing interviews with self-researchers, which we had also utilised for creating the personas [23]. From these interviews, we extracted a sample of excerpts pertaining to motivations, frustrations, and ideas related to their self-research practice and community engagement. These excerpts were shortened or paraphrased, then transferred to virtual post-its on an online collaboration board. The collected statements were subsequently clustered into overarching themes, as depicted in Figure 2.

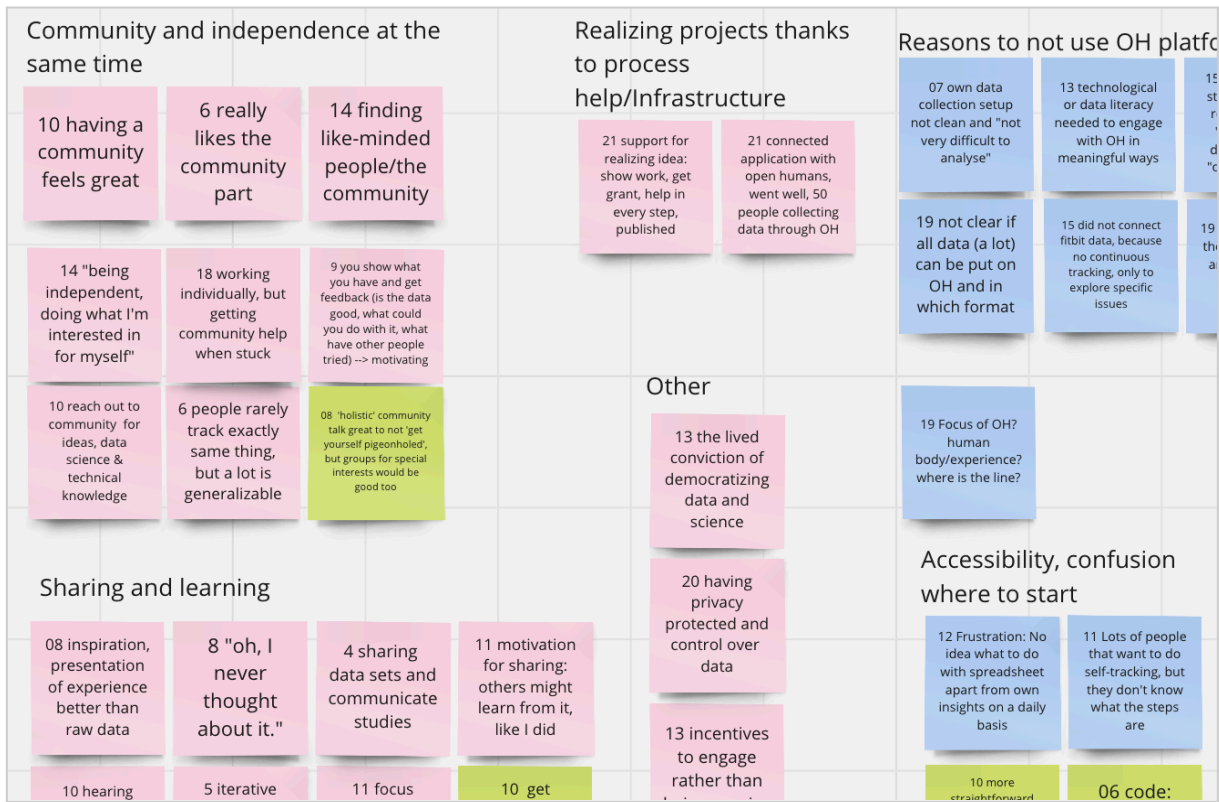

**Figure 2** *Screenshot of a part of the visual collaboration board, showing statements from interviews clustered by topic, and colour-coded according to the "rose-bud-thorn" method, pink: motivations, blue: frustrations, green: areas for improvement.*

Following the rose-bud-thorn method, each card was assigned a distinct colour, with pink denoting motivations, blue representing frustrations, and green indicating ideas or areas of potential. For instance, a motivation was expressed as "Finding like-minded people/a community," a frustration was articulated as "No time to attend community meetings," and an improvement idea was presented as "Make code less intimidating at first sight." We deliberated the statements, and classified them into prioritised areas for the ongoing design process, areas considered long-term goals that were not immediate for this design process, and deprioritized areas. The full list is in the supplementary materials. The process resulted in the identification of the following focus areas: "Something that encourages people to share their progress", "A way to find what has been done / the knowledge that already exists in the community in an unstructured way" in form of "some sort of cross-referenced information system", "Forming long-term social motivations", and "Analysis assistance".

## Prototyping

To start the ideation phase, a brainstorming session was conducted to complement the identified focus areas with potential approaches and anticipated challenges. For instance, the focus area "A way to find what has been done" was enriched with ideas for a cross-referenced information system, including features such as "For a given topic, what methods can you use? What worked versus what did not?" and "Make processes / tacit knowledge explicit." Anticipated issues were also outlined, such as "How is our approach different from forums or Reddit?" and "Problems with lists of tools: maintenance, spam, e.g., advertisement for apps from startups." Subsequently, a list of potential solution approaches was compiled, encompassing a custom personal science social platform, a community wiki, reusable programming notebooks or data management tools, and a reassessment of the Open Humans platform's usability to align user flows with personas.

The concept of a custom platform, considered the broadest in scope, was further developed into a mockup using Figma (see Supplementary Figure 1). This envisaged platform was conceptualised as a social network where users could document and discover projects, seek advice, and connect with individuals sharing similar interests. During a team discussion, various goals and motivations for using this platform were identified, such as sharing acquired knowledge, seeking assistance when stuck or in need of ideas for new projects, discovering what others have done, and connecting with and expanding the niche community. A key concern raised was the substantial development time and resources required upfront, with no guarantee of success and community adoption. Additionally, the concept heavily relied on user adoption and input to become valuable, particularly in motivating self-researchers to actively contribute.

Acknowledging these concerns, the team adapted the concept into a wiki format to enhance feasibility and set more realistic expectations for initial user engagement. While the original idea of a custom platform was not discarded, the wiki format was seen as a more achievable and practical approach to address the challenge of creating a cross-referenced information system.

## Personal Science Wiki

Following this prototyping, the Personal Science Wiki was implemented as a collaborative knowledge management system for self-researchers. This wiki is accessible at https://wiki.openhumans.org. MediaWiki was chosen as the underlying technical infrastructure due to its mature capabilities in supporting co-created knowledge management systems, its adaptability, and ease of deployment, allowing us to quickly gain user feedback from real usage. An essential feature is the open editing model, potentially minimising maintenance and moderation bottlenecks. Based on our brainstorming, we structured the initial information architecture around categories that are similar to those used in the archive of Quantified Self's Show and Tell talks, comprising Tools, Topics, Projects, and People (see Figure 3).

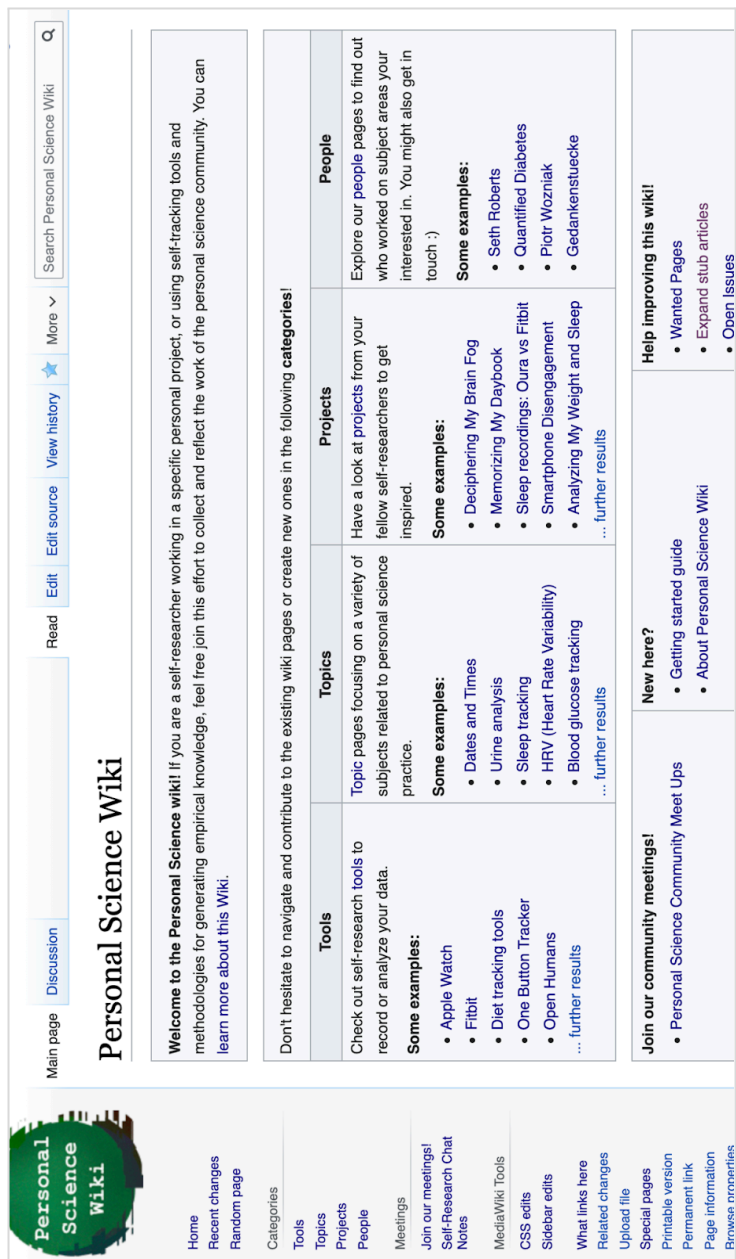

*Figure 3* An early version of the Personal Science Wiki homepage, showing the main category structure "Tools", "Topics", "Projects", and "People"

To facilitate automatic page linking, we integrated the "Semantic MediaWiki" [45] extension, an add-on for MediaWiki allowing the assignment of properties to pages. This feature enables semantic querying, allowing the aggregation of pages that share a common property. Templates were developed to utilise these properties, connecting content across main categories. In the "Infobox" template, users can assign other pages as properties. This allows, for instance, a page describing a self-tracking tool like the Oura Ring (see Figure 4) to be linked with associated topics such as sleep tracking or activity tracking. Additionally, we implemented a template which aggregates all pages linked back to the current page. This serves to showcase projects or self-researchers associated with a particular tool (c.f. Figure 4).

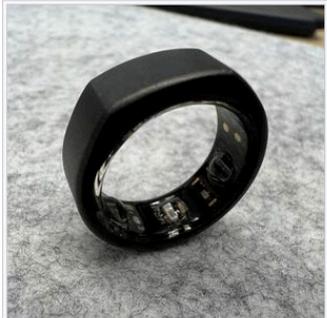

*Figure 4* An early version of the "Oura Ring" page, a page of the category "Tool", including a "Tool Infobox" template on the top right, and a "Linked content on this wiki" template on the bottom of the page

As a peer-produced platform reliant on user contributions, community engagement became a central focus and challenge. After developing a proof of concept, the platform was presented to a manager from Quantified Self who had not been previously involved in the design process to gauge its potential to fill a gap. Subsequently, a select group of community members was encouraged to use the early prototype of the wiki, providing valuable feedback to address some first usability issues. Various pathways were established to collect user feedback, including pages on the wiki for open issues and brainstorming ideas for improvement that any user could contribute to. Additionally, a dedicated #wiki channel was introduced in Open Humans' Slack, and segments of the weekly self-research community meetings were allocated for wiki discussions over several weeks. Subsequently, the wiki gained active usage within a small community circle. To seed content and connect a valuable, existing source of information within our system, we imported 373 documentations of Show and Tell talks from the Quantified Self archives as project pages. By April 2022, after six months, the wiki counted 476 content pages, thus 100 additional user-generated content pages on top of the imported ones, and 13 users with zero to nearly 700 edits. Notably, the most active contributor was external to our research team. In October 2022, during an online event on personal science, the wiki was officially launched to a broader audience. Between September 2021 and March 2023, the wiki featured in the notes of 56 weekly community meetings. These mentions ranged from discussing usability issues to general deliberations about the wiki approach, as well as links, when participants shared wiki pages they had edited or created.

During usage, some issues emerged. The integration of hundreds of content pages on self-research topics highlighted challenges with the existing category structure. Some pages did not align well with the main categories, leading to the "Topics" category becoming a somewhat unstructured catch-all for pages that did not fit elsewhere. Additionally, formal tests of intended user flows had not been conducted, and feedback had primarily come from a small group of core members and active wiki users. This prompted the planning of a user study to assess the wiki's usability and explore user-friendly information architectures for better structuring.

## User study

The study was run from 23 January to 24 March 2023, with 21 participants in total. All participants were prompted to do the card sorting activity, while 10 of the participants were also asked to engage with the usability study in the same session, following the card sorting. Five of these user tests were done in an initial set of usability tests, and another five in a second iteration after changes had been made based on the results of the first set. A description of the sample can be found in Table 4.

| **Study participants** | |
|---|---|
| Age | mean = 34.90 years; SD = 11.36; range = 19 – 66 |
| Gender | male = 12; female = 9; other = 0 |
| Education | PhD = 6; Master = 9, high school = 6 |
| Professional background | Engineering / research = 19 (predominantly from computer science = 9 and biology = 4); experience in health domain = 4 |
| Country of residence | Europe = 14; North America = 5; Eastern Asia = 1; Southern Asia = 1 |

*Table 4 Demographic information of the study sample*

The study also included questions about experience and interest in self-tracking, as well as wikis, including the Personal Science Wiki. The replies are summarised in Table 5.

| **Introductory questions regarding self-research** | |
|---|---|
| Self-tracking experience | none/almost none = 2; casual tracking = 11; intensive tracking/self-research projects = 9 |
| Personal science community involvement (Quantified Self/Open Humans) | Active involvement = 7; passive/casual involvement = 3; no involvement = 11 (of which 3 involved in other communities: weight monitoring forum, mental health Facebook groups, academic research in health monitoring) |
| Self-tracking areas of interest (multiple answers possible) | General fitness = 9 (physical activity, heart rate, blood oxygen); mental health = 6; sleep = 5; diet = 3; productivity = 3; chronic conditions = 2; meta research on personal science = 2; menstrual tracking = 1; location = 1; microbiome = 1 |

*Table 5 Study participants' self-reported involvement in self-research*

Motivations for participants to engage in self-tracking or self-research included monitoring, identifying patterns, pursuing specific goals such as weight loss, an interest in trying interventions and exploring the possibilities of technology to impact one's health, as well as meta-interests in personal science practice. To gain insights into participants' experiences and challenges with existing information sources, other questions targeted recent experiences and satisfaction with information resources related to personal science. Participants reported seeking information for various purposes, such as deciding on which tracking devices or software to purchase or download, learning how to use them, and gaining a general understanding of tracking topics and how to track specific variables. Participants turned to various sources, including online searches, scientific articles, blog articles, podcasts, Open Humans or Quantified Self forums, platforms like Reddit, advice from friends, and, more recently, artificial intelligence chatbots like ChatGPT or Bing Chat for personal science-related information. While seven participants expressed overall satisfaction, criticisms included concerns about the lack of trustworthy resources due to advertising, a shortage of unbiased information or research-backed systematic analyses, and conflicting opinions and reviews. Some participants also struggled to find practical information on certain topics. Notably, four participants identified a lack of information about personal science or self-research projects beyond simple monitoring. This included missing details on formulating research questions, interventions to address these questions, the connection between personal experiences and used methods, or lack of information about the practice at all.

## Usability test

In the initial round of usability tests, not all participants successfully completed all specific search tasks (see Supplementary Table 5). Some participants, even when successful, required additional time due to moments of hesitation and exploration of various paths. Incomplete or slow task completion was attributed to a lack of information scent on the homepage, such as the absence of keywords like "sleep," causing confusion about where to find sleep-related projects or tools. Additionally, links to the main category pages were often overlooked on the homepage, causing participants to miss opportunities to discover pages based on category tags. Subsequently, participants attempted to use search expressions like "projects sleep tracking" in the search bar, leading to sometimes unspecific search results due to the built-in mediawiki search function not being optimised for complex search terms. Another common issue was the difficulty in finding the section "Linked content on this wiki" as participants did not scroll down to the bottom of the page. Participants expressed that they did not expect important content below the 'reference' section.

In response, the interface of the homepage was rearranged to prominently display links to the main categories. The lists showing example pages were removed, as they misled some participants into thinking they represented all available content. Additionally, content showcasing community activities, an automatically updating calendar of upcoming events, and a featured article were added. A screenshot of the updated homepage can be found in Figure 5.

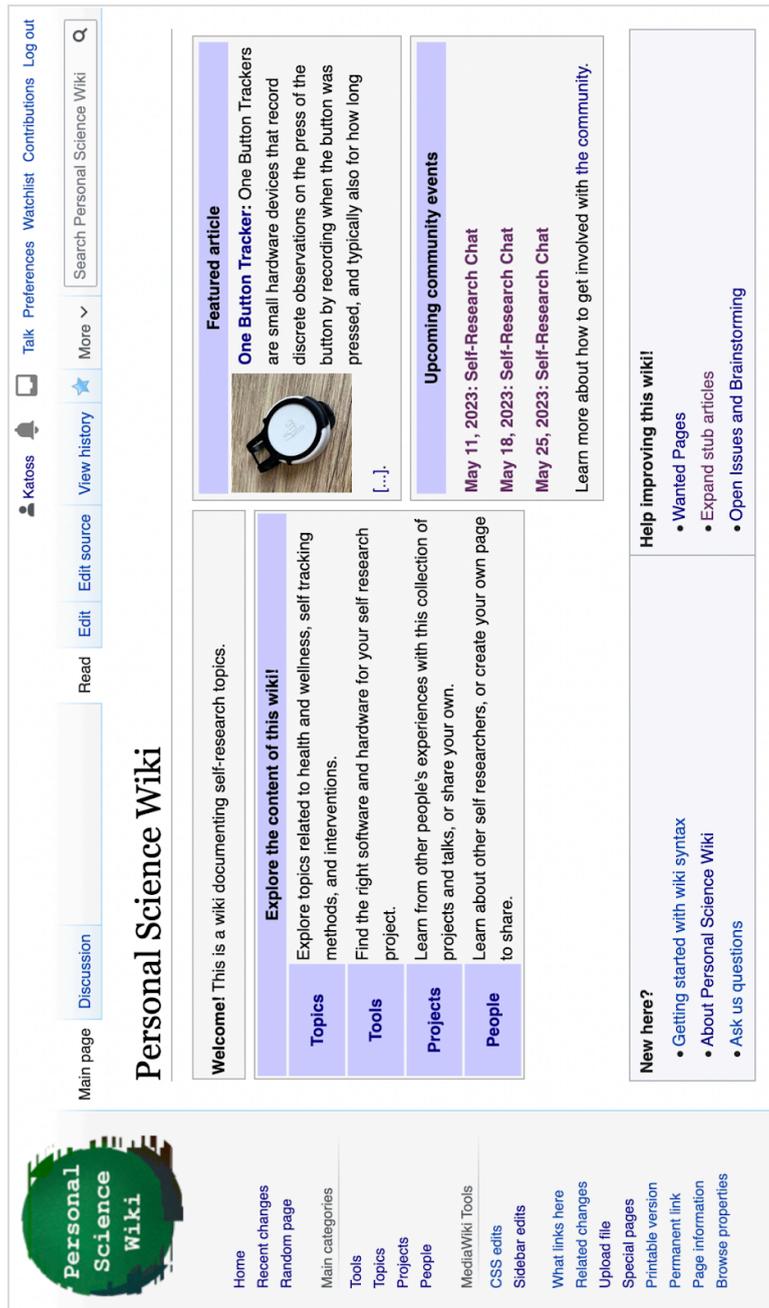

*Figure 5* Screenshot of the updated user interface of the Personal Science Wiki homepage after the first iteration of usability tests

Moreover, we made adjustments to the templates for the semantic properties on content pages. Recognizing that most participants did not find the "Linked content on this wiki" section at the bottom of the page, we relocated this section to the top right of the page within the infobox template. In this revised version, instead of displaying the entire list of connected pages, we opted to show only the count of linked pages for improved readability (see Figure 6). Users can access the detailed list with a click.

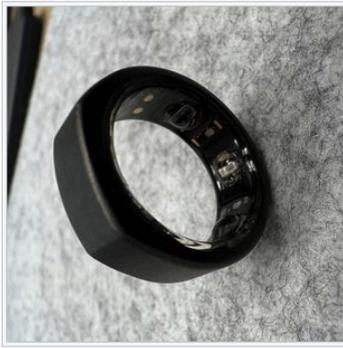

*Figure 6* Screenshot of the updated "Oura Ring" page, used as an example for a Tool page

A primary workflow the wiki aims to facilitate is the discovery of content, such as projects or tools, related to a specific topic of interest, like sleep or activity tracking. The key entry point for this is consulting the respective topic page and exploring linked content from there. To offer an alternative pathway to discover linked content, we introduced sub-categories to the main categories. The creation of sub-categories was informed by the semantic properties of existing pages, and by feedback from usability tests and early qualitative analysis of results from card sorting tasks. The new subcategories were then programmatically assigned to the pages using the MediaWiki API. For the "People" category, we opted not to add further subcategories due to the limited number of pages in the category, suggesting that labelling project pages by individuals might suffice. Table 6 provides a comprehensive list of all new subcategories.

| Top-level category | Subcategories |
|---|---|
| **Topics** | Blood testing and tracking tools; Body temperature tracking tools; Data analysis tools; Diet tracking tools; Fitness and heart rate tracking tools; Hardware; Mental health, journaling and self-report tools; Open source tools; Productivity, learning, and cognitive abilities tools; Sleep tracking tools; Software |
| **Tools** | Data analysis; Disease, pain and chronic condition; Discussions; Experiment design; Interventions; Personal Science Community; Things to track |
| **Projects** | Body measurement projects; Cognition and learning projects; Diet, digestion and weight loss projects; Disease, pain and chronic condition projects; Environment projects; Fitness and physical activity projects; Habits projects; Heart rate and cardiovascular health projects; How to's; Menstrual health, fertility and pregnancy projects; Mental health projects; Productivity projects; Sleep projects; Show and Tell (*note*: category existed before the usability test); Social life and social media projects |
| **People** | - |

*Table 6 Subcategories created on the wiki after the first iteration of usability tests*

After implementing these changes, the second iteration of usability tests saw an improvement in task completion for specific search tasks, rising from 66% to 100%. Almost all tasks were completed with little hesitation, apart from the task "Please find people who have worked on activity tracking". Four out of five participants hesitated upon realising that the "People" category lacked subcategories, unlike the other categories. Ultimately, they navigated to the aggregation through the "activity tracking" topic page, which all participants successfully located.

In the exploration tasks related to a sleep project and a topic of choice, additional insights into usability and user interests, needs, and expectations emerged. For the sleep tracking project, nine out of ten participants began by visiting the sleep tracking topic page and exploring linked content from there. One participant initially went to the Projects main category and consulted the sleep tracking projects subcategory. Participants visiting the sleep tracking topic page expressed the desire for an overview of the topic, including information about interventions, factors impacting sleep, and tracking devices. However, they lamented the lack of curated content on the page during the study.

Four participants found the list of linked sleep projects overwhelming (78 projects), struggling to determine which projects were the most relevant. The creative titles of project pages, imported from the Show and Tell talk archive (e.g. "Grandma was a lifelogger", "Sleep as a galaxy", were not perceived as helpful. Participants expressed a wish for a hierarchy or curated list. Regarding video recordings of project talks, participants wanted extracted information in a wiki-how style or a table containing details like methods, variables, goals, and results.

Nine out of ten participants repeated the task covering the following topics of interest: Apple Watch, blood glucose tracking/diabetes, note-taking and diaries, diet tracking, chronic pain, weight gain or loss, microbiome, and cognitive testing. All topics were present on the wiki, but some lacked extensive content. Participants appreciated detailed information about devices and sensors, including projects and specific details like nutrients. Missing information about tools prompted participants to express a desire to contribute. After the study, one participant added details to the Apple Watch page, and another participant added a new page for a note-taking tool.

One long-term community member provided general feedback, pondering the wiki's role amid emerging AI chatbots and alternative technologies like GitHub or notebooks. They saw the wiki as a potential centralised resource, stating:

*"I think there's a need for a centralised resource where people interested in doing experiments can find other experiments or experiments that are along that same line that they can modify appropriately for their own purposes. So like the immediate thing that you should be able to tell when you look at the personal science wiki is, you should be immediately able to find the experiments that kind of relate to the problem you're trying to solve."*

## Card sorting

All 21 participants engaged in the card sorting task. The datasets for analysis included transcripts of video recordings for qualitative analysis and tabular kardSort output for quantitative analysis. In line with best practices for cardsorting [41], data from one participant was excluded from the quantitative data due to creating a "miscellaneous" category that can bias the results. However, their comments were still considered for qualitative analysis.

Regarding the quantitative analysis, 20 participants created a total of 147 categories (mean = 8.4; SD = 1.8; range = 5 – 12). The results of the hierarchical clustering are depicted in a dendrogram in Figure 7. The colour cutoff was set at 80% or a distance of 16. This decision was based on perceiving clusters containing cards similar enough to represent a unique content category and distinct enough from other categories. This resulted in six categories with three to 14 cards, averaging eight cards each.

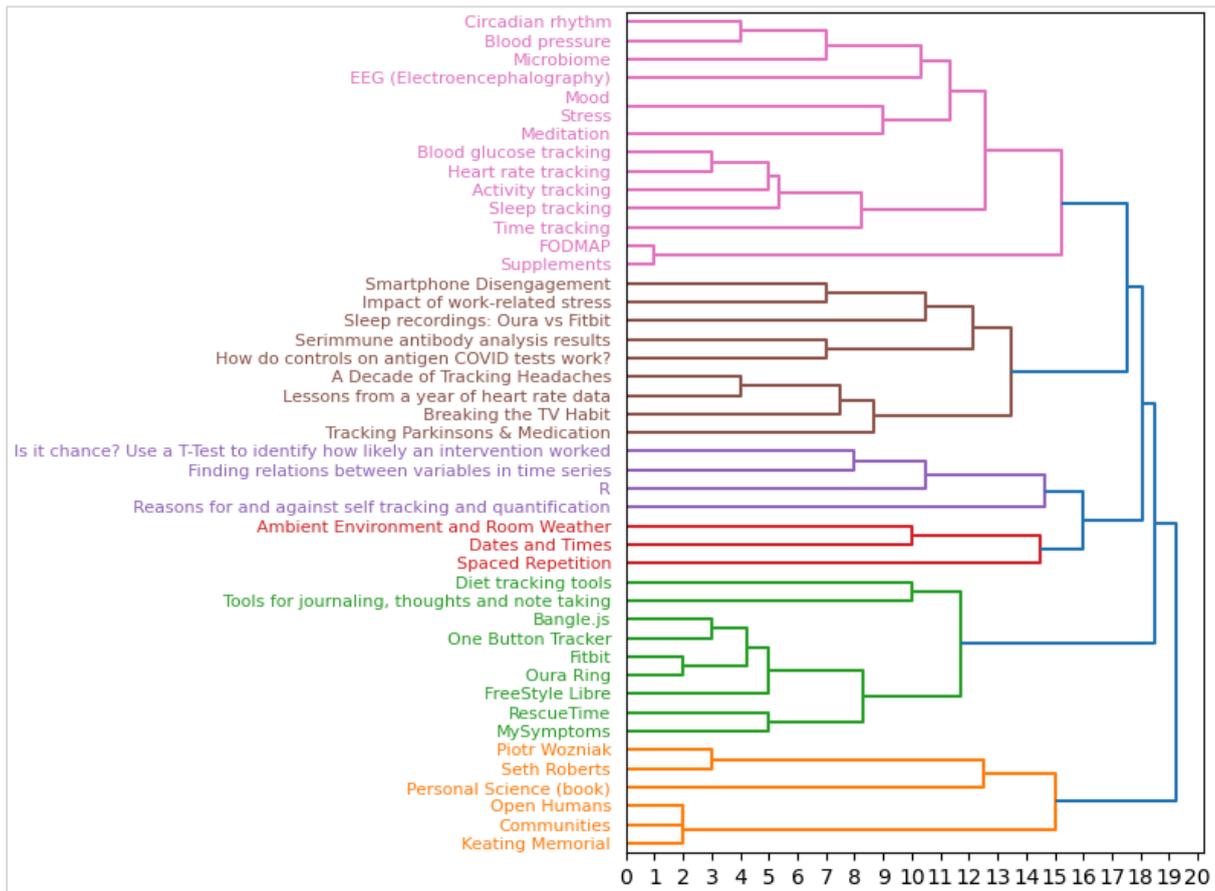

*Figure 7* Dendrogram visualising the results of hierarchical cluster analysis of the card sorting outputs of all 20 participants

As a next step, participant-generated labels were examined to comprehend the concepts covered in these clusters. Since hierarchical cluster analysis represents an average mental model, it is possible that no participant exactly replicated the clusters seen in the dendrogram. To still get an impression of participant-generated labels for the clusters, the following method was used: For each cluster, we first tried to see if there was a label assigned by a user for the whole set of cards. The same was repeated for subsets of cards in that cluster, removing one or more cards or subclusters. This allowed us to identify both concepts that roughly cover complete clusters, as well as concepts for subclusters in some cases. To identify concepts independent of exact wording, we clustered the labels based on similarity and tried to find overarching descriptions. A list of labels and sub-clusters can be found in the supplementary materials. Based on the results, the clusters (in chronological order) are described as follows:

- **Pink:** Labels are primarily related to tracking methods or variables to track (frequently related to the body); sub-clusters cover interventions or personal improvement, health topics, mental health, as well as diet and nutrition.
- **Brown:** Predominantly labelled as projects or experiments, but also as discussions or lessons learned, resources/media/Show and Tell talk, or community blog; sub-clusters cover mental health and specific health problems.
- **Violet:** Labels largely refer to methods, with a focus on data analysis, but also research questions or tools; other concepts that appear frequently are asked questions, general skills and self-tracking in general, studies, community, advice or blog, resources, and references.
- **Red:** Labels refer to parameters, variables, or entities, but also to methods; the card "Spaced repetition" also received labels regarding cognitive abilities or interventions.
- **Green:** Predominantly labelled as tools, including devices and apps, sometimes with keywords regarding measurement or tracking; another concept covered is tracking methods; sub-clusters regarding topics like general health and fitness or mental health.
- **Orange**: Predominantly labelled community, people, self-researchers; sometimes resources, references, or meta-information.

The qualitative analysis of the interview transcripts provided insights into the reasonings, difficulties, and organisational schemes used by individual participants. Some individual cards were occasionally criticised for covering unknown concepts (e.g., the "FODMAP" diet), being too broad to anticipate the content it represents on the wiki (e.g., "Stress"), or being challenging to place because they were perceived as a single instance of a potentially different category (e.g., the statistical programming language "R"). One participant questioned if some pages should be on the wiki at all, referring to general information on topics like the FODMAP diet, which they believed was more suitable for Wikipedia than a Personal Science Wiki.

Participants frequently expressed the desire to create hierarchies of categories, which is not possible with kardSort. For example, they suggested splitting the tools category into data collection and data analysis tools, software and hardware, or tools divided by variables they track. Another suggestion was to split projects by body part or health topic and divide general health and well-being topics into sub-categories regarding, e.g. chronic conditions. Participants frequently mentioned the coexistence of different organisational schemes, suggesting the possibility of sorting pages either by the type of resource (e.g., tool, project) or by subject matter (e.g., fitness, chronic condition). A participant with diabetes highlighted that a dedicated 'diabetes' category would be straightforward and useful for them, but topics like 'blood glucose tracking' could also be sorted into a different health tracking category for users with other interests. Participants also proposed different entry points to information based on users' interests and levels of experience. For example, they suggested a dedicated pathway for newcomers to guide them through foundational knowledge before delving into specific topics, as well as a shortened path for experts to quickly find specific information. Detailed results of the user study can be found in the supplementary materials.

# Discussion

Here, we describe how we used a design thinking approach to understand how the previously described barriers for individuals to successfully engage in personal science could be addressed. Moving into the implementation, our work made use of participatory design principles to investigate how technical infrastructure for the peer-production of knowledge can help overcome the challenges of scaling up the community and the tendency of individuals to initiate projects from scratch.

## Identifying barriers to personal science practice

Early in our work, we defined the particular type of scaling we aimed to address in our design process, focusing on facilitating scalability to enable more individuals to participate in the entire breadth of personal science practice, from questioning to discovery. While other scaling approaches, such as automating common processes and data donation, were considered, they were ultimately discarded. Automating common processes, centred on providing ready-to-use tools to eliminate technical barriers, had been explored by the Open Humans platform, offering data import connectors, uploaders for various data sources, and a Jupyter notebook integration [22]. However, this approach was dismissed due to potential challenges in maintaining it at a larger scale, given the frequent changes in APIs or tools. Similarly, for data donation, where individuals contribute personal data to a collective dataset for scientific studies [46], Open Humans already provides tools.

Next, we identified barriers faced by individuals who want to engage in personal science, and recognized the importance of skill-building. Addressing personal health questions demands a comprehensive set of research skills, including developing research questions, selecting tools, data collection, analysis, interpretation, and method adaptation to individual contexts [47]. Individuals frequently rely on peer support to get directions and acquire skills, and belonging to and sharing experiences with a community was often cited as a strong motivator for continuous engagement in the practice. However, we identified some issues with the current community knowledge sharing practices: Many individuals do not have time to attend regular community meetings and these meetings are not easily scalable to accommodate a large number of participants for discussing individual projects. While asynchronous, text-based alternatives like forums, blogs, social media, or chats exist, these thread-based approaches have limited capabilities for creating consensus knowledge, often remaining as a collection of individual approaches. Attempts to create overviews and best practices encounter maintenance issues and quickly become outdated, not reaching the status of a shared community knowledge base [16]. The popular narrative format of Show & Tell talks faces challenges in revealing failed attempts and iterations, as the narrative makes it seem as if the process had been streamlined from the start. Finally, the user study highlighted people's frustration with the lack of reliable online resources, encountering conflicting opinions and advertising when seeking self-tracking information. These findings provide insight into possible reasons why self-researchers tend to create their methods from scratch as they undertake projects.

## Addressing the barriers with a wiki approach

After some design iterations, we opted for a wiki-based approach to provide an infrastructure for a shared community knowledge base. Wikis facilitate the creation of consensus knowledge by allowing multiple users to edit pages and engage in discussions through attached "Talk" pages. Anyone can edit any page, alleviating maintenance bottlenecks by not depending on specific individuals. Entry barriers are low, with no mandatory user account creation and the flexibility to perform minimal edits or create entire pages. Early contributors can benefit from the wiki as a space to document and share projects without depending on input from others. Moreover, wiki approaches have been chosen in similar contexts such as chronic pain management [48] or DIY research projects in environmental justice [49]. Wikis offer a quick and easy setup, facilitating the gathering of real user feedback on a functional prototype. This aligns with the recommended approach in tool design, prioritising practical user experience over abstract discussions [50]. Wikis are an instance of peer production, a form of distributed knowledge production relying on self-organising communities of individuals, notably known from Wikipedia and open-source software development [51]. This approach has been emphasised as beneficial for open-ended creative tasks, and for attracting a diverse group of intrinsically motivated individuals with different skills and motivations [52]. This aligns with the nature of the task, the development of personal research projects, with the user types we prioritised in the beginning of the design phase, as well as with values within the personal science community, such as peer support, sharing, transparency, and empowerment. The Personal Science Wiki is thought to complement existing platforms, as a permanent, open, community-owned space for curating and preserving content, connecting pages in a web of semantic links.

## Lessons on information needs and representation

The usability test served a dual purpose by addressing and resolving wiki infrastructure usability issues while uncovering broader user requirements for personal science information resources. During the open exploration task, participants demonstrated a realistic understanding of the wiki's content scope, choosing topics within its existing coverage. This suggested that even at its current scale, the wiki covers common areas of interest in health and well-being tracking.

Personal science often involves tacit knowledge – skills and know-how acquired through practice [53], the transfer of which is a common knowledge management challenge [54]. Participants frequently expressed the wish for an entry point providing an overview of each tracking topic, covering common variables or tracking tools, best practices or interventions. They also wondered about curated lists of a few exemplary projects for each topic, caused by the overwhelming nature of the existing, automatically aggregated project lists based on tags. While participants appreciated the richness of project information, they criticised the current tagging system for lacking clarity in distinguishing projects specifically focused on an aspect (e.g., sleep), or those that just track that aspect as one variable of many. The titles of Show and Tell projects were deemed too open to quickly understand what the project was about. Standardisation could be a means to enhance clarity. Moverover, the Show and Tell format, suitable for event contexts, faced challenges when integrated into the wiki. Participants desired summary tables or manuals for quick access to relevant information within these pages. Since all of these pages can be edited in a wiki, features like summary tables can be added retrospectively, creating a direction for future work. Lastly, lack of specific pieces of information regarding their topics of interest motivated some participants to add this information to the wiki themselves during or after the study. This highlights the potential of the low entry barriers of wikis to increase casual contributions and improve low quality contributions, as outlined by [55].

## Lessons on information architectures

The card sorting study provided valuable insights into users' mental models regarding how they organise health-related self-research information, offering lessons for structuring knowledge resources like the Personal Science Wiki. While the initial category system

(Tools, Topics, Projects, People) roughly represented mental models, participants provided input for improvement. The findings suggest that the broad "Topics" category could be split into a category covering "health topics" or "tracking topics" as well as methods, especially regarding data analysis, interventions, and community-related resources. Participants also expressed a desire to introduce hierarchies, suggesting potential subcategories like, for example, general fitness and chronic conditions for health topics, hardware and software for tools, and physical and psychological for variables. Another lesson was that organisational schemes varied not only between individuals but also within individuals. Participants often considered cards to belong to multiple categories, reflecting the flexibility and complexity inherent in organising information. Two main organisational schemes emerged: ordering pages by "health topic," such as diabetes or heart rate tracking, and by "type of resource," more similar to the original category structure. Quantitative analysis favoured the organisation by the type of resource, with emerging subcategories for health topics. Importantly, the study suggested the use of complementary models, recognizing that a single, one-size-fits-all solution might not exist. This is in line with recommendations from the W3C consortium, suggesting the provision of multiple pathways to create robust navigation experiences [56]. Participants emphasised the importance of flexible organisational schemes, allowing users to choose based on their interests and skill levels, suggesting the need for entry points for newcomers offering overviews of general information and shortcuts for experts. Another factor to consider is that with a growing content base, requirements to the information architecture might change. This underscores an advantage of the wiki approach, which allows users to modify and create new categories, fostering the development of "folksonomies" [57] – community-created organisation themes that can evolve with the growth of the resource.

## Limitations and future work

One limitation of this study is the startup challenge faced by the Personal Science Wiki, a common issue in many peer production projects: Generating content requires buy-in from a sufficient number of contributors, while having enough content is crucial for attracting and retaining those contributors [58]. However, there are promising indicators for existing use cases and the adoption of the wiki, despite the currently small user base. Over two years, there has been active and continued engagement from a small circle of contributors, with links to pages frequently shared in community meetings. The wiki has occasionally been recommended as a knowledge resource for personal science on other social platforms. Furthermore, issues with the Show and Tell talk format surfaced during integration and testing in the wiki, pointing towards potential future research directions. Future work should focus on extracting relevant keywords and procedural information from these talks. Additionally, field studies involving individuals planning self-research projects can provide deeper insights into information needs outside of a theoretical, lab context, as seen in the user study. Finally, the study results can serve to refine and test new information architectures for personal science information.

Apart from that, the wiki's co-creation and testing were conducted with a specific target group – highly educated individuals predominantly residing in the global north, with a professional background in engineering and research. While this is representative of typical profiles engaging in this practice [59,60], it does not provide insights into the requirements and accessibility of the wiki for other potential user groups. Future research should consider the requirements of different demographics, enhancing accessibility for a more diverse audience interested in personal health inquiry. Finally, at the current scale, governance, moderation, and efforts to prevent misinformation were minimal. However, potential growth and the inclusion of new target groups may pose challenges that need careful consideration in the future.

## Conclusion

Using a design thinking approach, we identified that peer support can offer individuals help in overcoming barriers to practising personal science, including the lack of systematic access to community knowledge. To address this, we engaged in participatory design and co-designed a wiki-based infrastructure as a shared knowledge base, aimed at alleviating maintenance bottlenecks and enabling community-created project documentation and consensus knowledge.

Through our user study, we discovered that individuals expect overviews and curated information about tracking and intervention tools, approaches, and projects, with the documentation of self-tracking projects being a unique selling point of the wiki. We found that the frequently used Show and Tell talk format for project documentation poses some challenges when transferred to a knowledge management system. Issues include choosing informative titles and tags to aid in identifying relevant projects, as well as a lack of means to quickly extract information regarding research questions, methods, and outcomes.

The topics present on the wiki align with user expectations, but depth of content is still in need of improvement through more extensive user contributions. Regarding information architectures, our results indicate benefits of using multiple organisation schemes, notably ordered by resource type and tracking or health topic area, to facilitate discovery and error recovery. Similarly, providing different entry points for users with varying levels of expertise and search use cases, as well as utilising category hierarchies, can enhance information scent. Future research efforts are needed to explore how these knowledge organisation challenges of self-researchers' information needs can be overcome, in particular prioritising underrepresented demographics to broaden access to personal health inquiry.

# Data accessibility statement

Data and supplementary materials have been archived within the Zenodo repository: https://doi.org/10.5281/zenodo.10659150 [61].

# Ethics statement

Ethical approval for this user testing was obtained from the Institutional Review Board of the French National Institute of Health and Medical Research (Inserm, IRB00003888) on January 10, 2023.

# Authors' contributions

BGT: conceptualization, funding acquisition, investigation, methodology, project administration, software, supervision, validation, writing - original draft, writing – review & editing; GIW: data curation, resources, validation, writing – review & editing; KK: conceptualization, data curation, formal analysis, investigation, methodology, project administration, software, visualisation, writing – original draft; MPB: conceptualization, investigation, methodology, software, supervision, validation, writing – review & editing; SJ: data curation, resources, validation, writing – review & editing

All authors gave final approval for publication and agreed to be held accountable for the work performed therein.

# Use of Artificial Intelligence (AI) and AI-assisted technologies

AI-tools (Google Docs auto-complete and ChatGPT 3.5) were only used to improve readability and language of this work.

# Conflict of interest declaration

MPB is the executive director, and BGT the director of research of the Open Humans Foundation. GIW is the co-founder, and SJ an editor/data engineer of Quantified Self Labs.

# Funding

Thanks to the Bettencourt Schueller Foundation's long-term partnership, this work was partly supported by the Center for Research and Interdisciplinarity Research Fellowship (awarded to BGT).

# Acknowledgements

The authors would like to thank Enric Senabre Hidalgo for continued support throughout the project and providing insights and data on challenges faced by personal science practitioners. We would like to thank Ame Elliott, Benjamin Mako Hill, Vineet Pandey, Mélanie Dulong de Rosnay, Katarzyna Wac, and Molly Claire Wilson for providing support in the advisory and thesis defence committees of Katharina Kloppenborg. We want to thank Gary Wolf and Steven Jonas, all participants of the Open Humans and Quantified Self communities for their input and support, as well as the study participants.